\documentclass[aps,pre,twocolumn,amsmath,amssymb,superscriptaddress]{revtex4-2}
\usepackage[colorlinks]{hyperref}
\usepackage{graphicx}
\usepackage[dvipsnames]{xcolor}
\usepackage{amsmath}
\usepackage{braket}
\usepackage{soul}
\usepackage{color}
\usepackage{amsfonts}
\usepackage{ulem}

\newcommand{\im}{{\rm i}}

\newcommand{\vtheta}{ \pmb{\vartheta}}

\definecolor{darkgreen}{rgb}{0.20,1,0.0}
\definecolor{blue1}{rgb}{0 0.4470 0.7410}
\definecolor{green1}{rgb}{0.4660 0.6740 0.1880}
\definecolor{green2}{rgb}{0.3,0.9,0.3}
\definecolor{red1}{rgb}{0.8500 0.3250 0.0980}
\definecolor{yellow1}{rgb}{0.9290 0.6940 0.1250}
\definecolor{purple1}{rgb}{0.4940 0.1840 0.5560}
\definecolor{grey1}{rgb}{0.6 0.6 0.6}
\definecolor{black1}{rgb}{0.3 0.3 0.2}
\date{\today}

\newcommand{\sutdphys}{Science, Mathematics and Technology Cluster, Singapore University of Technology and Design, 8 Somapah Road, 487372 Singapore}
\newcommand{\sutdepd}{EPD Pillar, Singapore University of Technology and Design, 8 Somapah Road, 487372 Singapore}

\newcommand{\cqt}{Centre for Quantum Technologies, National University of Singapore 117543, Singapore} 
\newcommand{\majulab}{MajuLab, CNRS-UNS-NUS-NTU International Joint Research Unit, UMI 3654, Singapore}  
\newcommand{\sichuan}{College of Physics and Electronic Engineering, and Center for Computational Sciences, Sichuan Normal University, Chengdu 610068, China}

\newcommand{\R}{\textcolor{black}}

\begin{document}
	
\title{Continuous-time parametrization of neural quantum states for quantum dynamics}  
 
\author{Dingzu Wang}   
\affiliation{\sutdphys}
\affiliation{\cqt}
\author{Wenxuan Zhang}
\affiliation{\sutdphys}
\affiliation{\cqt} 
\author{Xiansong Xu} 
\affiliation{\sichuan}  
\affiliation{\sutdphys} 
\author{Dario Poletti} 
\affiliation{\sutdphys}
\affiliation{\sutdepd}
\affiliation{\cqt}
\affiliation{\majulab}

\begin{abstract}  
Neural quantum states are a promising framework for simulating many-body quantum dynamics, as they can represent states with volume-law entanglement. As time evolves, the neural network parameters are typically optimized at discrete time steps to approximate the wave function at each point in time. Given the differentiability of the wave function stemming from the Schr\"odinger equation, here we impose a time-continuous and differentiable parameterization of the neural network by expressing its parameters as linear combinations of temporal basis functions with trainable, time-independent coefficients. We test this ansatz, referred to as the smooth neural quantum state (\textit{s}-NQS) with a loss function defined over an extended time interval, under a sudden quench of a non-integrable many-body quantum spin chain. We demonstrate accurate time evolution using a restricted Boltzmann machine as the instantaneous neural network architecture. \R{We show} that the parameterization \R{enables accurate simulations with fewer variational parameters, independent of time-step resolution. Furthermore,} the smooth neural quantum state \R{also} allows us to initialize and evaluate the wave function at times not included in the training set, both within and beyond the training interval.
\end{abstract}

\date{\today}

\maketitle 

\section{Introduction} 
The study of the dynamics of quantum many-body systems plays a central role in understanding fundamental aspects of quantum statistical mechanics \cite{DAlessioRigol2016}. Driven by recent experimental and theoretical progress, research in this area has rapidly advanced, enabling unprecedented access to coherent real-time evolution and revealing a range of nonequilibrium phenomena, including prethermalization \cite{MoeckelKehrein2008, GringSchmiedmayer2012, MoriUeda2018, ReimannDabelow2019, MallayyaRigol2021}, many-body localization \cite{BASKO20061126, abanin2019colloquium}, dynamical quantum phase transitions \cite{heyl2013dynamical, heyl2018dynamical}, many-body scars \cite{turner2018weak}, and slow relaxation regimes \cite{CapizziPoletti2025}. Despite these advances, simulating such dynamics remains a major challenge in computational quantum physics, primarily due to the exponential growth of the Hilbert space with system size, which renders exact approaches intractable. Tensor network methods provide an efficient representation for area-law entangled states in one-dimensional systems and have achieved notable success, but they struggle to accurately simulate higher-dimensional systems or states with complex entanglement \cite{verstraete2008, SchollwockMPS, orus2014}.

Neural quantum states (NQS), which represent quantum wave functions using artificial neural networks, have recently emerged as a flexible variational ansatz capable of describing volume-law states, even in dimensions larger than one \cite{DengDasSarma2017, carleo2019machine, Lange_2024}. 
Over the past few years, NQS have been successfully applied to compute time evolution of many-body quantum systems, both for unitary~\cite{2024NC,2023PRXQ4d0302M,schmitt2020quantum, gutierrez2022real, SinibaldiVicentini2023, donatella2023dynamics, burau2021unitary, SchmittZurek2022, ZhangPoletti2024, GravinaVicentini2024, vandewalle2024manybodydynamicsexplicitlytimedependent, sinibaldi2024timedependentneuralgalerkinmethod,2025arXiv250310462C, 2021arXiv210802200W} and dissipative~\cite{vicentini2019variational, nagy2019variational, yoshioka2019constructing, hartmann2019neural, reh2021time} systems. 
Several approaches have been developed to evolve the networks in time, including time-dependent variational Monte Carlo (tVMC)\cite{carleo2017solving}, its projective extension (p-tVMC)\cite{sinibaldi2023unbiasing}, and more recently proposed global loss-based optimization schemes~\cite{sinibaldi2024timedependentneuralgalerkinmethod, vandewalle2024manybodydynamicsexplicitlytimedependent}.
A review of the current state of the art is provided in Ref.~\cite{schmitt2025simulatingdynamicscorrelatedmatter}, and important technical details concerning sampling strategies and loss functions can be found in Ref.~\cite{GravinaVicentini2024}. Meanwhile, Ref.~\cite{ZhangPoletti2024} presents an approach to incorporating stochastic reconfiguration \cite{Sorella1998, sorella2007weak} within the p-tVMC framework for time evolution.
However, notwithstanding these developments, reliably capturing the dynamics of quantum many-body systems using NQS remains highly nontrivial.
 
In isolated quantum systems, the Schrödinger equation induces a unitary and continuous evolution of the quantum state in Hilbert space, suggesting that the dynamics is amenable to variational representations that are continuous and differentiable in time. 
Motivated by this insight, we propose a continuous and differentiable parametrization in time of the neural network ansatz, which we refer to as \textit{smooth} neural quantum state (\textit{s-}NQS) ansatz (see Fig.~\ref{fig:1}). 
Specifically, each time-evolving parameter $\vartheta_j(t)$ in the \textit{s-}NQS is defined as a linear combination of smooth temporal functions, e.g., polynomials, with time-independent variational parameters $ \{\theta_{j,q}\} $. 
This parametrization allows for the network to accurately represent quantum states not only at discrete training times but also at unoptimized points within the same time interval $ \tau $  through interpolation (see Fig.~\ref{fig:1}). Moreover, by extending the temporal functions beyond the training interval, the \textit{s}-NQS can simulate quantum states at later times.
This feature provides a well-informed initialization for the subsequent interval, which can accelerate convergence and significantly enhance training stability. 

For each interval, the parameters of the \textit{s}-NQS are determined by smooth temporal functions and variational parameters. This parameterization enables the definition of a global loss function over each time interval, which may reduce error accumulation compared to step-by-step schemes.
Note that the use of global loss functions and time-dependent networks, has also been explored in recent studies~\cite{sinibaldi2024timedependentneuralgalerkinmethod, vandewalle2024manybodydynamicsexplicitlytimedependent}, showing promising results. However, our method employs a distinct loss formulation and a different network parameterization. We demonstrate the effectiveness of our method by simulating quantum quenches—where the system is initially prepared in the ground state of one Hamiltonian and subsequently evolved under a different one as done, for example, in studies of thermalization~\cite{kollath2007quench, polkovnikov2011colloquium, yukalov2011equilibration, cassidy2011generalized, canovi2011quantum, rigol2009quantum}. 

The paper is organized as follows: in Sec.~\ref{sec:network}, we introduce the \textit{smooth} neural quantum state ansatz, the $s$-NQS; in Sec.~\ref{sec:optimization}, we describe the optimization strategy for the $s$-NQS, from the loss function to the gradient computation; in Sec.~\ref{sec:quench} we show our results for the Hamiltonian quench considered, and draw our conclusions in Sec.~\ref{sec:conclusions}. More details about the method and the calculations can be found in App.~\ref{app:window}-\ref{app:taylor}.

\section{The $s$-NQS ansatz}\label{sec:network} 

\begin{figure}[tpb]
    \centering
    \includegraphics[width=0.85\linewidth]{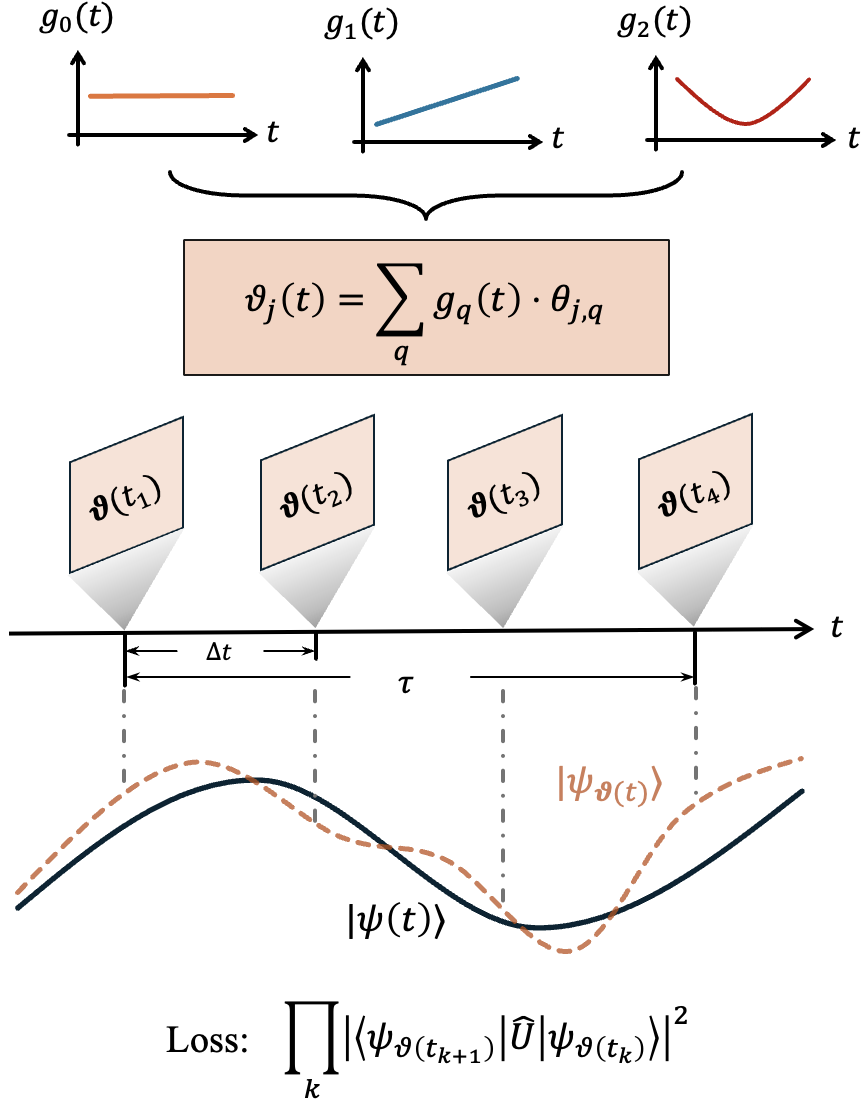}
    \caption{ Schematic illustration of the \textit{s}-NQS approach.    Top: the temporal basis functions $\{g_q(t)\}$ are linearly combined with time-independent coefficients $\{\theta_{j,q}\}$ to construct the network parameter $ \vartheta_j(t) $.
    Middle: these parameters determine the variational wave function $ |\psi_{\boldsymbol{\vartheta}(t)}\rangle $ (brown dashed line), within a time interval $ \tau $ composed of smaller steps $ \Delta t $. 
    Bottom: the parameters are optimized by minimizing a global loss function that enforces consistency with the exact evolution state $ |\psi(t)\rangle $ (black solid line) over the entire interval, using the time evolution operator $\hat{U}$.
    }
    \label{fig:1}
\end{figure} 

In this section, we introduce the \textit{smooth} neural quantum state ansatz. 
Within this approach, a wave function at time $t$, $|\psi(t)\rangle$, is approximated by $|\psi_{\boldsymbol{\vartheta}(t)}\rangle$ using a set of time-evolving network parameters $\vtheta(t) = \{ \vartheta_1(t), \dots, \vartheta_{j}(t), \cdots, \vartheta_{N_p}(t) \}$, where $N_{p}$ denotes the total number of parameters, and each $\vartheta_j(t)$ is defined as a linear combination of smooth temporal basis functions $\{g_q(t)\}$ and a set of variational parameters $\{\theta_{j,q}\}$ 
\begin{align}
    \vartheta_j(t) = \sum_{q=0}^{Q-1} g_{q}(t) \cdot \theta_{j,q}, \label{eq:vtheta}
\end{align}
where $ Q $ is the number of temporal basis functions. This construction ensures that the temporal behavior of the network parameters is intrinsically smooth, a property that is encoded directly into the parameterization, as schematically illustrated in Fig.~\ref{fig:1}. The derivative  variational parameters can be written as
\begin{align}
    \frac{\partial}{\partial \theta_{j,q}} 
    = \frac{\partial}{\partial \vartheta_{j}} \frac{\partial \vartheta_j}{\partial \theta_{j,q}} 
    = g_{q}(t) \frac{\partial}{\partial \vartheta_{j}}. \label{eq:dirivative}
\end{align}
This allows us to compute derivatives with respect to $\theta_{j,q}$ at a given time. 
In this work, we use shifted Chebyshev polynomials $\mathcal{T}_q$ as the temporal basis functions $g_q(t)$,
\begin{align}
    g_{q}(t) = \mathcal{T}_q(r(t)), \label{eq:chebyshev}  
\end{align}
where $\mathcal{T}_q$ denotes the $q$-th Chebyshev polynomial and $r(t)$ rescales the time variable such that $r(t) \in [-1, 1]$ (see App.~\ref{app:window} for details). The choice of temporal basis functions could influence the expressivity and stability of the \textit{s}-NQS, and we leave a detailed investigation of this aspect for future work.

To specify the wave function representation within the \textit{s}-NQS framework, we adopt the restricted Boltzmann machine (RBM) architecture, which was introduced in Ref.~\cite{carleo2017solving} as the first neural-network-based ansatz for quantum states. The RBM is a shallow, fully connected neural network and allows for an explicit and analytic expression of the wave function.
For a quantum spin system with $L$ sites, the RBM maps each spin configuration $\boldsymbol{x} = (x_1,\dots,x_L)$, where $x_i \in \{-1, +1\}$, to an unnormalized wave function $\psi_{\boldsymbol{\vartheta}(t)}(\boldsymbol{x})=\langle\boldsymbol{x}|\psi_{\boldsymbol{\vartheta}(t)}\rangle \in \mathbb{C}$, expressed as      
\begin{align}
    \psi_{\boldsymbol{\vartheta}(t)}(\boldsymbol{x}) 
    =& 
    \exp \left( \sum_{i = 1}^{L} a_{i}(t) x_{i}  \right) \nonumber \\
    &\times\prod_{j = 1}^{M} 2 \cosh \left( b_{j}(t) + \sum_{i = 1}^{L} W_{i,j}(t) x_{i} \right),
\end{align}
where $\boldsymbol{\vartheta}(t) = \{\boldsymbol{a}(t), \boldsymbol{b}(t), \boldsymbol{W}(t)\}$ represents a set of complex-valued parameters of the RBM: $a_i(t)$ and $b_j(t)$ are visible and hidden biases, respectively, and $W_{i,j}(t)$ are elements of the weight matrix. The visible layer takes the spin configuration as input, while the hidden layer captures nontrivial correlations via the non-linear activation. We set the number of hidden units to $M = \alpha L$, where $\alpha \in \mathbb{Z}^+$ determines the number of hidden units and affects the expressive capacity of the network. All network parameters are time-dependent following the smooth parametrization defined in Eq.~\eqref{eq:vtheta} and collectively grouped into the vector $\boldsymbol{\vartheta}(t)$. Owing to its analytic structure, the RBM allows for closed-form evaluation of the wave function and logarithmic derivatives with respect to network parameters.

\section{Variational Optimization Scheme}\label{sec:optimization}

We consider the unitary time evolution of a quantum state governed by the Schrödinger equation (working in units such that $\hbar = 1$),
\begin{align}
    \frac{d}{dt} |\psi(t)\rangle = -\mathrm{i} \hat{H} |\psi(t)\rangle.
\label{eq:schrodinger}
\end{align}
The solution is given by the application of the time evolution operator $\hat{U}(t)$ to the initial state $ \left| \psi(0) \right\rangle  $,
\begin{align}
    |\psi(t)\rangle = \hat{U}(t) |\psi(0)\rangle,
\end{align}
where $ \hat{U}(t) = \exp(-\im \hat{H} t) $.

In numerical simulations, we discretize time using a small {\it time step} $\Delta t$, which allows for an approximate implementation of the evolution operator. This is particularly relevant for large systems, where exact evaluation of $\hat{U}(t)$ is typically intractable and approximate methods are required. Further details regarding the approximation of $\hat{U}$ will be presented in Sec.~\ref{sec:taylor_expansion} and App.~\ref{app:taylor}. 
To manage the evolution over longer time ranges, we introduce two additional temporal scales that structure our optimization procedure. The first is the {\it time interval} $\tau$, which consists of several consecutive time steps and defines the domain over which the variational parameters of the \textit{s}-NQS ansatz are optimized globally. The second is the {\it time window} $T$, which comprises multiple such intervals and sets the support of the temporal basis functions $\{g_q(t)\}$. 
In the following, we introduce the loss function used to train the \textit{s}-NQS within each interval, describe how its gradients are evaluated, present the approximation of the evolution operator, and explain how the long-time evolution is assembled across successive time windows.

\subsection{Global loss function over time intervals}\label{sec:loss_function} 

To determine the variational parameters of the \textit{s}-NQS within each time interval, we define a global loss function that quantifies the fidelity between the network state and its time-evolved counterpart. The loss is defined as
\begin{align}
\mathcal{C}(\{\theta_{j,q}\})
&=
\prod_{k = 0}^{K} \mathcal{C}_{k}(\{\theta_{j,q}\}),
\label{eq:loss}
\end{align}
where $K$ is the number of time steps $\Delta t$ within the interval $\tau$, and $\mathcal{C}_k$ quantifies the fidelity at each step. For $k \neq 0$, we define
\begin{align} 
\mathcal{C}_{k} (\{\theta_{j,q}\})
&=
\frac{ \langle \psi_{\boldsymbol{\vartheta}(t'')} \vert \hat{U} \vert \psi_{\boldsymbol{\vartheta}(t')} \rangle }{\langle \psi_{\boldsymbol{\vartheta}(t'')} \vert \psi_{\boldsymbol{\vartheta}(t'')} \rangle }
\frac{\langle \psi_{\boldsymbol{\vartheta}(t')} \vert \hat{U}^\dagger \vert \psi_{\boldsymbol{\vartheta}(t'')} \rangle}{\langle \psi_{\boldsymbol{\vartheta}(t')} \vert \psi_{\boldsymbol{\vartheta}(t')} \rangle},
\end{align}
with $t' = t_0 + (k - 1) \Delta t$ and $t'' = t_0 + k \Delta t$ and, for brevity, we denote $\hat{U}(\Delta t)$ simply as $\hat{U}$. Each $\mathcal{C}_k$ corresponds to the normalized overlap between the variational states before and after time evolution. 
This loss function correctly reflects Hilbert-space distance between quantum states and is structurally similar to the one used in the projected time-dependent variational Monte Carlo (p-tVMC) approach~\cite{SinibaldiVicentini2023, ZhangPoletti2024, GravinaVicentini2024}. 
At the initial time step $k = 0$, we include the initial state $\ket{\psi(t_0)}$ in the loss function as
\begin{align}
\mathcal{C}_{0} (\{\theta_{j,q}\})
&=
\frac{\langle \psi_{\boldsymbol{\vartheta}(t_0)} \vert \psi(t_0) \rangle }{\langle \psi_{\boldsymbol{\vartheta}(t_0)} \vert \psi_{\boldsymbol{\vartheta}(t_0)} \rangle}
\frac{\langle \psi(t_0) \vert \psi_{\boldsymbol{\vartheta}(t_0)} \rangle }{\langle \psi(t_0) \vert \psi(t_0) \rangle},
\end{align}
where $\ket{\psi(t_0)}$ corresponds to the known representation of wave function at the beginning of the interval. We then maximize the loss function $\mathcal{C}$ to optimally determine the \textit{s}-NQS representation of the time-evolved states across the interval. 

The structure of the \textit{s}-NQS ansatz plays a key role in the optimization process. Specifically, the smooth temporal functions $ \{g_{q}(t)\} $ naturally capture the continuity within and across intervals.
In practice, each network parameter at time $t_c$ within the current interval can be initialized from the previous interval 
\begin{align}
    \vartheta_j(t_{c}) = \sum_{q} g_{q}(t_{c}) \cdot \theta^{[t_0 - \tau,\;t_0]}_{j,q}, 
\end{align}
where we used the notation $\theta^{[t_0 - \tau,\;t_0]}_{j,q}$ to specify that these variational parameters have been optimized in the previous time interval $[t_0 - \tau,t_0]$, while $t_c$ is in the current time interval $t_c \in [t_0, t_0 + \tau]$. 
Owing to the smooth parametrization, the resulting network states—evaluated using parameters from the previous interval—are typically already close to the optimal representation in the current one, as we will show later.
As a result, it accelerates convergence and improves the overall stability of the training process.

Another key point is the definition of the $\mathcal{C}_0$ term in the global loss function [Eq.~\eqref{eq:loss}]. 
Here, $\mathcal{C}_0$ does not involve the time-evolution operator $\hat{U}$ and instead directly measures the overlap between the variational state and the known initial state.
This alignment between initialization and loss structure further enhances the stability and efficiency of the optimization process. 

\subsection{Gradient evaluation}\label{sec:gradients}

The variational parameters $\theta_{j,q}$ are optimized over each time interval $\tau$ using gradient descent      
\begin{align}
    \theta_{j,q} \gets \theta_{j,q} - \eta \frac{\partial \mathcal{C}}{\partial \theta_{j,q}},
\end{align}
where $\eta$ is the learning rate. \R{While this update rule provides a general form, we use the AdamW optimizer~\cite{loshchilov2018decoupled} in practice, which adaptively adjusts learning rates and includes weight decay regularization.} To evaluate the gradient of the global loss function $\mathcal{C}$ under the smooth parametrization of the \textit{s}-NQS, we use the chain rule to obtain
\begin{align}
    \frac{\partial \mathcal{C}}{\partial \theta_{j,q}} 
    &= 
    \sum_{k=0}^K \left(\prod_{k' \ne k} \mathcal{C}_{k'}\right) 
    \frac{\partial \mathcal{C}_k } {\partial \theta_{j,q}}  \nonumber \\
    &=  
    \sum_{k=0}^K \left(\prod_{k' \ne k} \mathcal{C}_{k'}\right) \frac{\partial \mathcal{C}_k } {\partial \vartheta_{j}} \cdot g_q(t_k), \label{eq:gradient}
\end{align}
where $t_k = t_0 + k \Delta t$. This expression follows from the smooth parameterization in Eq.~\eqref{eq:vtheta}, where each $\theta_{j,q}$ contributes to the full time-dependent parameter $\vartheta_j(t)$ through the smooth temporal function $g_q(t)$.

The value of each $\mathcal{C}_k$ for $k \neq 0$ is given by
\begin{align} \label{eq_overlap}
    \mathcal{C}_k 
    &=
    \left[\sum_{\pmb{x}} P_{t''}(\pmb{x}) \mathcal{C}_\text{loc}^{t''}(\boldsymbol{x})\right]
    \left[\sum_{\boldsymbol{y}} P_{t'}(\pmb{y}) \mathcal{C}_\text{loc}^{t'}(\boldsymbol{y})\right],
\end{align}
where the local estimators are defined as
\begin{align}
    \mathcal{C}^{t''}_\text{loc}(\boldsymbol{x})
    &= 
    \sum_{\boldsymbol{x}'} \frac{\psi_{\boldsymbol{\vartheta}(t')}(\boldsymbol{x}')}{\psi_{\boldsymbol{\vartheta}(t'')}(\boldsymbol{x})} \langle \boldsymbol{x} \vert \hat{U} \vert \boldsymbol{x}' \rangle, \\
    \mathcal{C}^{t'}_\text{loc}(\boldsymbol{y})
    &= 
    \sum_{\boldsymbol{y}'} \frac{\psi_{\boldsymbol{\vartheta}(t'')}(\boldsymbol{y}')}{\psi_{\boldsymbol{\vartheta}(t')}(\boldsymbol{y})} \langle \boldsymbol{y} \vert \hat{U}^{\dagger} \vert \boldsymbol{y}' \rangle.
\end{align}
For $k = 0$, the time-evolution operators reduce to the identity, resulting in Kronecker deltas $\delta(\boldsymbol{x}, \boldsymbol{x}')$ and $\delta(\boldsymbol{y}, \boldsymbol{y}')$ in the local estimators.
The probabilities are given by
\begin{align}
    P_{t''}(\pmb{x}) 
    = 
    \frac{|\psi_{\boldsymbol{\vartheta}(t'')}(\boldsymbol{x})|^{2}}{\langle \psi_{\boldsymbol{\vartheta}(t'')} \vert \psi_{\boldsymbol{\vartheta}(t'')} \rangle }, \;
    P_{t'}(\pmb{y})
    =
    \frac{|\psi_{\boldsymbol{\vartheta}(t')}(\boldsymbol{y})|^{2}}{\langle \psi_{\boldsymbol{\vartheta}(t')} \vert \psi_{\boldsymbol{\vartheta}(t')} \rangle }.
\end{align}
The Hilbert space dimension grows exponentially with system size. Therefore, we employ Monte Carlo (MC) sampling to evaluate both the local estimators and the associated probabilities for large systems.
\R{To this end, we perform Markov Chain Monte Carlo (MCMC) sampling using the Metropolis-Hastings algorithm with single-spin flip proposals applied at randomly chosen sites. Throughout this work, we draw $N_{\text{s}} = 5000$ samples per optimization step from the probability distribution $ P_{t}(\pmb{x})$. To reduce autocorrelation, we save samples every $ 10 \times N $ Metropolis updates, where $ N $ is the system size.}
The gradients $\partial \mathcal{C}_k / \partial \vartheta_j$ can be computed either analytically or using automatic differentiation. In this work, \R{we use the AdamW optimizer~\cite{loshchilov2018decoupled} with an initial learning rate of $10^{ - 3} $, weight decay $ 10^{ - 5} $, and AMSGrad enabled~\cite{pytorch}. The learning rate is adaptively reduced during training using the ReduceLROnPlateau scheduler with patience $400$, cooldown $40$, a decay factor of $0.5$, and a minimum learning rate of $10^{ - 6}$~\cite{pytorch}. This dynamic scheduling helps balance training stability and convergence.}

\subsection{Approximate evaluation of the evolution operator}\label{sec:taylor_expansion}  

In large many-body systems, the exact application of the evolution operator $ \hat{U} = \mathrm{e}^{-\im \hat{H} \Delta t} $ is computationally intractable, as $\hat{U}$ is a dense operator whose action on a basis state involves exponentially many nonzero terms.
To overcome this challenge, several approximation strategies have been proposed, including Trotter-Suzuki decomposition that splits the evolution into local terms Ref. ~\cite{2007CMaPh270359B, 2005LNP67937H}, Taylor expansions around small time steps, and multiplicative \R{product schemes such as the Taylor-root expansion and Padé product expansion~\cite{2024NC,GravinaVicentini2024}.} 
Here, we adopt a second-order Taylor expansion of the form
\begin{align}
    \hat{U}(\Delta t) 
    \approx
    \hat{\mathbb{I}} -\im \Delta t \hat{H} - \frac{1}{2} (\Delta t \hat{H})^{2},
\end{align}
where $\hat{\mathbb{I}}$ denotes the identity operator and $\Delta t$ is assumed to be small enough such that higher-order terms can be neglected. This approximation yields a simple analytic form for $\hat{U}$ and enables efficient Monte Carlo sampling in high-dimensional Hilbert spaces. 
\R{While higher-order Taylor expansions remain fully compatible with our framework and can be employed when higher accuracy is required, multiplicative product schemes which decompose the propagator as a product of operator terms of the form $ \hat{U} = \prod_{k} (\hat{\mathbb{I}} + c_{k}\mathrm{i}\Delta t \hat{H}) $ offer linear step-complexity and thus improved computational scaling at high orders.} More details can be found in App.~\ref{app:taylor}. 

\subsection{Continuity across time windows}\label{sec:windows} 

In this context, the smooth temporal basis functions $\{g_q(t)\}$ are defined over a rescaled time domain $r(t) \in [-1, 1]$, which corresponds to a finite segment of physical time referred to as a time window $T$.
As a consequence, the smooth parameterization of the \textit{s}-NQS is inherently confined to this window.
To extend the simulation beyond a single time window while preserving temporal flexibility, we assemble the full time evolution by sequentially joining multiple windows.

A key advantage of the \textit{s}-NQS framework is that it provides not only the values of the network parameters at any given time, but also their time derivatives.
By evaluating those values at the window boundary, the \textit{s}-NQS enables smooth continuation into the next time window.  For example, in the first time window from $ t = 0 $ to $t = T$, the network parameters are given by
\begin{align}
    \vartheta_j(t)
    &= 
    \sum_{q = 0}^{Q-1} \mathcal{T}_{q}(r_1(t)) \cdot \theta^{[0,\;T]}_{j,q}, 
\end{align}
where $r_1(t)$ denotes the rescaled time in the first window and $\theta_{j,q}^{[0,\;T]}$ means the variational parameters are optimized in the time window $[0,\;T]$. In the subsequent window, we define a new parameterization
\begin{align}
    {\vartheta}_j(t)
    &= 
    \sum_{q = 0}^{Q-1} \mathcal{T}_{q}(r_2(t)) \cdot \theta^{[T,\;2T]}_{j,q}
\end{align}
At the interface between two time windows, we enforce continuity of all time derivatives of the network parameters. This leads to the matching condition
\begin{align} \label{eq:connectingwindows} 
    \!\!\!\!\!\!\!\!\!\!\!\!\!\!\!\frac{d^{n} }{d t^{n}} &\left[\sum_{q} \mathcal{T}_q(r_1(t)) \cdot \theta^{[0,\;T]}_{j,q} \right]_{r_1(t)=1} \nonumber \\ 
    &\;\;\;\;\;\;\;\;\;\;= \frac{d^{n} }{d t^{n}} 
\left[ \sum_{q'} \mathcal{T}_{q'}(r_2(t)) \cdot \theta^{[T,\;2T]}_{j,q'} \right]_{r_2(t)=-1}\!\!\!\!\!\!\!\!\!\!\!\!\!\!\!\!\!\!\!\!\!\!. 
\end{align} 
This matching condition ensures the continuity of the $ n $-th time derivatives of the network parameters across time windows. More details can be found in App.~\ref{app:window}. 

\section{Simulation of Hamiltonian quench}\label{sec:quench} 

To demonstrate the effectiveness of the \textit{s}-NQS approach in simulating quantum many-body dynamics, we consider the tilted Ising model (TIM) with open boundary conditions. The Hamiltonian for a system with $L$ spins is given by
\begin{align}
\hat{H}
=
J \sum_{i = 1}^{L - 1} \hat{\sigma}_{i}^{z} \hat{\sigma}_{i + 1}^{z}
- \sum_{i = 1}^{L} (h_{x} \hat{\sigma}_{i}^{x} + h_{z} \hat{\sigma}_{i}^{z}),
\end{align}
where $J$ is the nearest-neighbor coupling constant, $h_x$ and $h_z$ denote external magnetic fields along the $x$- and $z$-directions respectively, and $\hat{\sigma}_i^{a}$ (with $a = x,y,z$) are Pauli operators acting on site $i$.
This system is known to be interacting and nonintegrable unless one of the fields $ h_{x} $ or $ h_{z} $ is zero, and it has been extensively used to study eigenstate thermalization hypothesis~\cite{PhysRevA432046, PhysRevE50888} and many-body quantum chaos~\cite{2016AdPhy65239D}. 
When $J$ dominates over $h_x$, the ground state is either ferromagnetic or antiferromagnetic. Conversely, when $ h_{x} $ dominates, the ground state is paramagnetic. In the case of $ h_{z}=0 $, the model reduces to the integrable transverse field Ising model which, in 1D, exhibits a quantum phase transition at the critical point $J = h_{x}$.

In our simulations, we initialize the system in paramagnetic product state of the form
\begin{align}
    \ket{\psi(0)} 
    &= \bigotimes_{i=1}^L \frac{1}{\sqrt{2}} \left( \ket{0}_i + \ket{1}_i \right),
\end{align}
where $\ket{0}_i$ and $\ket{1}_i$ denote the eigenstates of $\sigma_i^z$ on site $i$.
This state corresponds to the ground state of the Hamiltonian in the limit $h_x \gg J $ and $ h_{z} $, where transverse field dominates and the system is deep in the paramagnetic phase. We then quench the Hamiltonian parameters into a regime in which the ground state exhibits antiferromagnetic correlations.

\subsection{Benchmarking on a fully computable small system}
\begin{figure}[tpb]
    \centering
    \includegraphics[width=1.\linewidth]{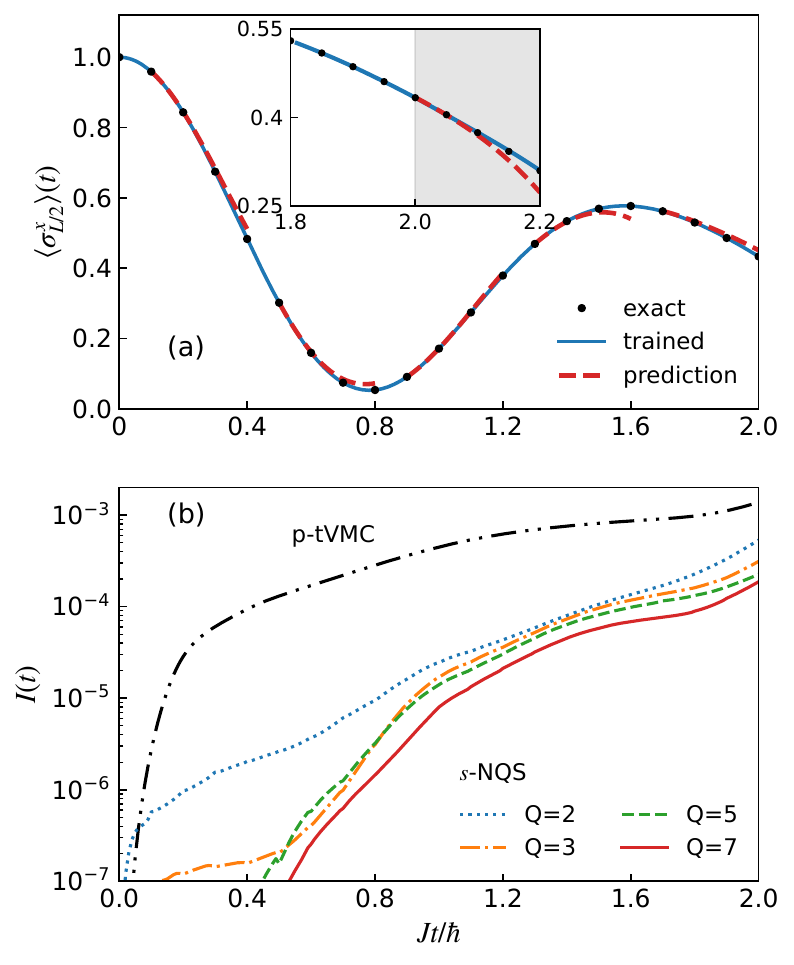}
    \caption{
    Quench dynamics of 1D tilted Ising model with open boundary conditions. (a) Time evolution of the transverse magnetization of the middle site $\langle \sigma^x_{L/2} \rangle$; (b) Time evolution of the infidelity between numerical solutions and the exact results. We compare the performance of p-tVMC with $s$-NQS for different values of the temporal expressivity parameter $Q$. Parameters are $\alpha = 5$, $Jdt/\hbar = 0.01$, $L = 10$, $h_x = h_z = 0.3J$ and $J\tau/\hbar= 0.1$.   } 
    \label{fig:2}
\end{figure} 
To evaluate the intrinsic performance of the \textit{s}-NQS ansatz, we first focus on a small spin system where all quantities can be computed exactly without approximations. Specifically, we consider a spin chain of length $ L = 10 $, for which all observables and gradients can be evaluated exactly by summing over the full basis of $ 2^{10} = 1024$ configurations, and the evolution operator $ \hat{U} = \exp(-\im \hat{H} \Delta t) $ can be directly exponentiated. We choose Hamiltonian parameters $h_x = h_z = 0.3J$, and use a time step $J \Delta t = 0.01$ with time intervals $J \tau = 0.1$. The evolution is performed up to $Jt = 2.2$, while the time window is set to $JT = 2.0$, which implies that we need to connect two different windows.

Fig.~\ref{fig:2}(a) shows the time evolution of the transverse magnetization at the center of the chain, $\langle \sigma^x_{L/2}\rangle(t)$. 
Exact results are represented by black dots, while the solid blue line denotes the \textit{s}-NQS simulation using the temporal basis size $ Q = 3 $ and a RBM with $\alpha=5$. 
The excellent agreement shows the accuracy of the ansatz when trained within a time interval. 

A key strength of the \textit{s}-NQS lies in its smooth parameterization, which naturally provides high-quality initializations for adjacent, yet-to-be-trained intervals. This is illustrated in Fig.~\ref{fig:2}(a) by the red dashed lines, which show the network prediction in multiple untrained intervals using only the parameters optimized in the preceding interval. For instance, the first dashed line (from $t = \tau$ to $t = 4\tau$) is obtained directly from training over the interval $t \in [0, \tau]$, without further optimization. Similarly, training from $t = 4\tau$ to $5\tau$ enables prediction from $5\tau$ to $8\tau$. These results demonstrate that the smooth structure already provides near-optimal initialization, enabling accurate predictions even before additional training is performed. 

The inset of Fig.~\ref{fig:2}(a) highlights the performance of the \textit{s}-NQS when bridging across adjacent time windows. The first time window ends at $t = T = 2/J$, after which the parameters for the subsequent window are initialized using the continuity conditions defined in Eq.~\eqref{eq:connectingwindows}. Without any additional training in the second window, the resulting red dashed line accurately reproduces the observable's evolution up to $t \approx 2.1/J$, demonstrating the strength of the smooth initialization. Upon further optimization within the new window, the final trained trajectory (solid blue line) aligns precisely with the exact result, as indicated by the black dots.

In Fig.~\ref{fig:2}(b), we show the infidelity of the simulated wave functions as a function of time for different values of $Q$ in the $s$-NQS. The infidelity is defined as $I(t)=1-|\langle \psi_{\rm exact}(t)| \psi_{\pmb{\vartheta}(t)}\rangle|^2$ where $|\psi_{\rm exact}(t)\rangle$ is obtained via exact diagonalization. Given the size of the system considered, the infidelity can be evaluated exactly. A larger value of $Q$ allows the wave function to capture finer time variations, effectively fitting more points within a given interval.
This increases the expressibility of the s-NQS. However, it comes at the cost of a larger number of parameters.
In Fig.~\ref{fig:2}(b), we also plot the infidelity obtained using p-tVMC \cite{sinibaldi2023unbiasing, ZhangPoletti2024}. 
For the latter, we used AdamW as an optimizer for a direct comparison to our $s$-NQS approach.  
For both the p-tVMC and $s$-NQS approaches, we use a RBM with $\alpha=5$ as the base neural network. 
All expectation values and gradients are computed exactly, without sampling. 
We observe that increasing $Q$ from $2$ to $7$ leads to a substantial reduction in infidelity. 
Interestingly, the \textit{s}-NQS achieves an infidelity that is orders of magnitude lower than that of p-tVMC. 

It is also important to compare the total number of parameters needed to perform these evolutions.
For the p-tVMC evolution over $K$ time steps, we require $K$ neural networks, and consequently $K N_p$ parameters, where $N_p$ is the number of neural network parameters at each time. This would allow us to recompute all observables at any of the time steps $k \Delta t$ \footnote{\R{If one chooses a limited set of observables a priori} and does not wish to be able to recompute the evolution over certain time steps, one can keep only 2$N_p$ parameters.}. In contrast, the \textit{s}-NQS requires only $QN_p$ parameters, and since $Q<K$ this leads to a significant reduction in memory usage.
Furthermore, the wave function can be evaluated at arbitrary time points, not limited to $k \Delta t$, and remains accurate even beyond the training interval. 

\R{\subsection{Effect of integration time step on the performance of \textit{s}-NQS}}

\begin{figure}[tpb]
    \centering
    \includegraphics[width=0.94\linewidth]{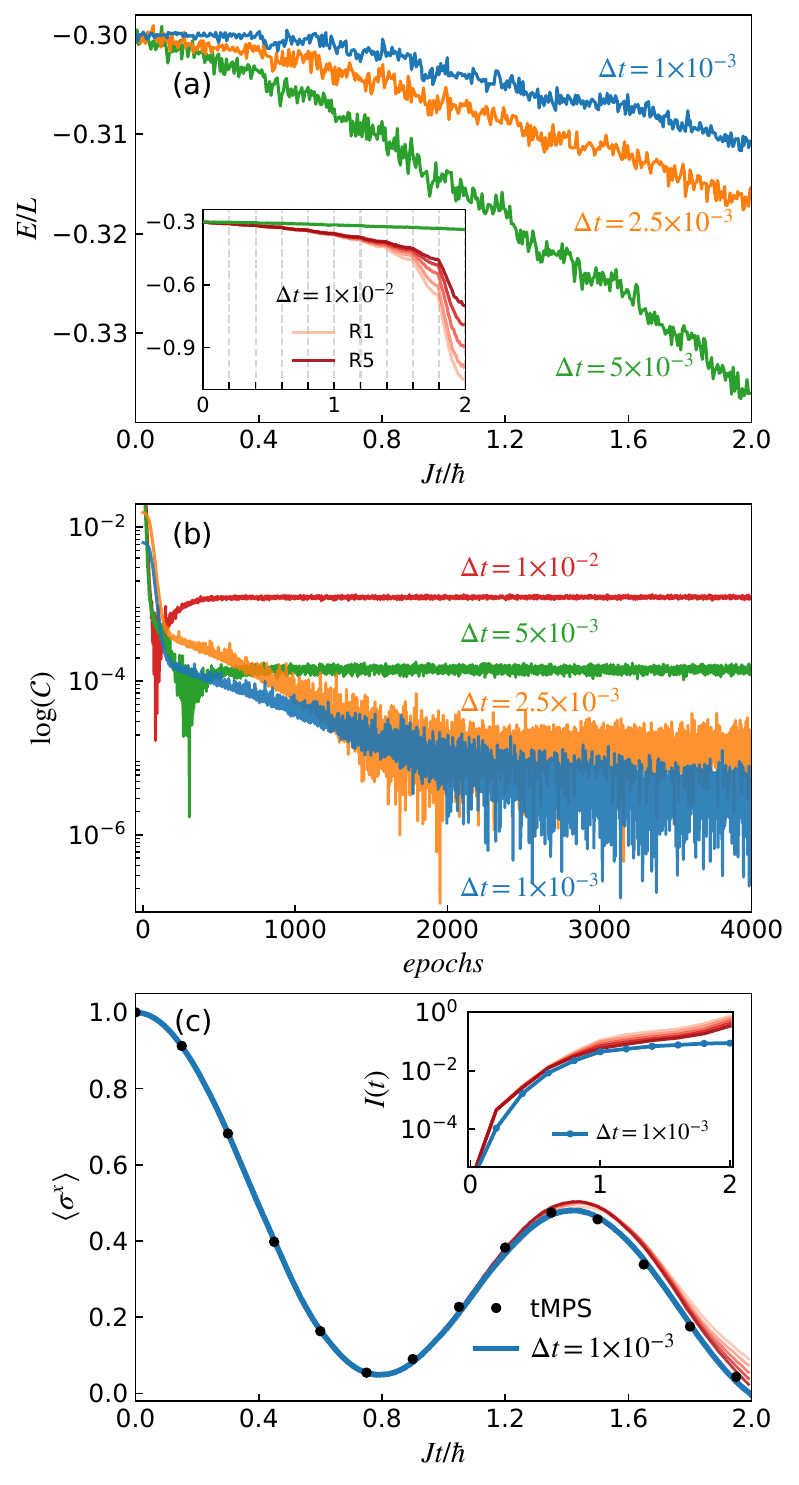}
    \caption{
        \R{
        Simulation of a quantum quench in the 1D tilted Ising model with $L = 30$ spins and open boundary conditions. (a) Time evolution of the energy density $E/L$ for different time steps $ \Delta t $. 
        Inset: effect of successive training rounds at $\Delta t = 10^{-2}$, shown as gradient red curves labeled R1 (light) to R5 (dark). 
        (b) Evolution of the logarithmic loss function $\log(\mathcal{C})$ over training epoch within the first interval. 
        (c) Time evolution of the transverse magnetization $\langle \sigma^x \rangle$, with \textit{s}-NQS results (curves) compared to tMPS benchmarks (black dots). Red curves correspond to five training rounds at fixed $\Delta t = 10^{-2}$, shown with increasing color intensity; the blue curve shows the final result at reduced $\Delta t = 10^{-3}$. The inset shows the infidelity estimated via Monte Carlo sampling with $50{,}000$ samples, quantifying the overlap between the \textit{s}-NQS and tMPS wavefunctions. 
        The parameters are $Q = 5, \alpha = 5, J \tau/\hbar = 0.2$, and $JT/\hbar = 2$.
        } }
    \label{fig:3}
\end{figure}
\R{In larger system simulations, the choice of the integration time step $\Delta t$ plays a crucial role in ensuring both numerical stability and physical accuracy. In this section, we perform a systematic study of how $\Delta t$ affects the performance of the \textit{s}-NQS method.} 
We \R{consider} a system \R{of} $ L = 30 $ spins \R{with} open boundary conditions. 
The system parameters remain the same as in the small-system case, and we use a time interval $J \tau = 0.2 $, and a time window $ JT = 2 $. 
\R{The parameters of the \textit{s}-NQS ansatz are set to $ Q = 5 $ and $ \alpha = 5 $.}

\R{To explore the influence of $ \Delta t $, we perform multiple training rounds at successively smaller time steps. In this procedure, parameters optimized at one resolution are used to initialize training at the next.} \R{This} coarse-to-fine strategy \R{is naturally supported by the parameterization of the \textit{s}-NQS ansatz, }whose time-independent parameters $\{\theta_{j,q}\}$ allow for predictions at untrained intermediate times.

\R{The evolution of the energy density $E/L$ during the training rounds is shown in Fig.~\ref{fig:3}(a). 
We first fix the time step at $\Delta t = 0.01$ and perform five training rounds (shown as the gradient red curves (R1 to R5) in the inset of Fig.~\ref{fig:3}(a)), each initialized from the previous one.
This stepwise optimization helps the ansatz better approximate the time-evolved states at this relatively coarse resolution. 
The inset of Fig.~\ref{fig:3}(a) shows that while significant energy drift occurs at this resolution, the deviation is gradually reduced over successive rounds.
After this stage, we progressively reduce the time step to $\Delta t = 5 \times 10^{-3}, 2.5 \times 10^{-3},$ and $10^{-3}$, using the optimized parameters from the previous step as initialization. 
When $\Delta t$ is decreased to $5 \times 10^{-3}$ (green line), even this moderate refinement in time resolution yields a marked reduction in energy drift. 
As $\Delta t$ is further reduced to $2.5 \times 10^{-3}$ and $10^{-3}$, the energy conservation continues to improve with the drift shrinking steadily. 
These results show the critical role of $\Delta t$ in determining the accuracy of the \textit{s}-NQS. 
Smaller time steps help reduce the accumulation of truncation errors introduced by the second-order Taylor expansion used to approximate the time-evolution operator. These errors, including deviations from unitarity, can hinder training stability during global optimization.}

\R{Fig.~\ref{fig:3}(b) shows the logarithmic loss function $\log(\mathcal{C})$ over training epochs in the first time interval $Jt/\hbar \in [0, 0.2]$. We find a clear trend that smaller time steps lead to lower values decreasing from approximately $ \mathcal{O}(10^{-3}) $ at $\Delta t = 10^{-2}$ to $ \mathcal{O}(10^{- 6}) $ at $\Delta t = 10^{-3}$. As smaller $\Delta t$ results in more time steps, the decrease in total $\log(\mathcal{C})$ implies an even greater improvement in loss at the level of individual time points.
Notably, for larger time steps, the behavior of $\log(\mathcal{C})$ initially decreases but subsequently rises again and plateaus, suggesting limited progress in later stages of training.
This rise-after-fall pattern is progressively delayed to later epochs as $\Delta t$ is reduced, and disappears entirely at $\Delta t = 0.001$.}
\R{This behavior may be linked to cumulative truncation errors in the second-order Taylor expansion and could be mitigated by more accurate propagation schemes, such as higher-order or product-based schemes~\cite{2024NC,GravinaVicentini2024}.}

The time evolution of $\langle \sigma^x\rangle$ is shown in Fig.~\ref{fig:3}(c). 
The results obtained from \textit{s}-NQS (curves) are shown in comparison with those from time-dependent matrix product states tMPS ~\cite{SchollwockMPS, verstraete2008, 2003PhRvL91n7902V,2004PhRvL93d0502V, WhiteFeguin2004, 2004JSMTE04005D} (black dots) simulations. 
\R{The red curves represent \textit{s}-NQS simulations with $\Delta t = 10^{-2}$ across multiple training rounds, showing gradual improvement. The blue curve ($\Delta t = 10^{-3}$) exhibits agreement with the tMPS results over the entire interval $Jt/\hbar \leq 2$.}
\R{These results demonstrate that smaller time steps significantly improve the accuracy of observables obtained from the \textit{s}-NQS.}
\R{To further assess wavefunction accuracy beyond observables, the inset of Fig.~\ref{fig:3}(c) displays the infidelity $1 - |\langle \psi_{\text{tMPS}}(t) | \psi_{\text{\textit{s}-NQS}}(t) \rangle|^2$ estimated via Monte Carlo sampling ($50{,}000$ samples).
The gradient red curves correspond to five rounds at fixed $\Delta t = 10^{-2}$ and the blue curve shows the result at the $\Delta t = 10^{-3}$.
These results demonstrate that smaller time steps improve accuracy. For $\Delta t = 10^{-3}$ the infidelity remains below $10^{-1}$ over the entire evolution window, confirming fairly close agreement with the tMPS benchmark.}

\R{Notably, all results are obtained using a fixed number of variational parameters ($\alpha = 5$, $Q = 5$). This demonstrates the efficiency of the \textit{s}-NQS parameterization in capturing time evolution at increasingly finer resolutions without increasing model complexity.}

\subsection{Larger system simulation}

\begin{figure}[tpb]
    \centering
    \includegraphics[width=0.95\linewidth]{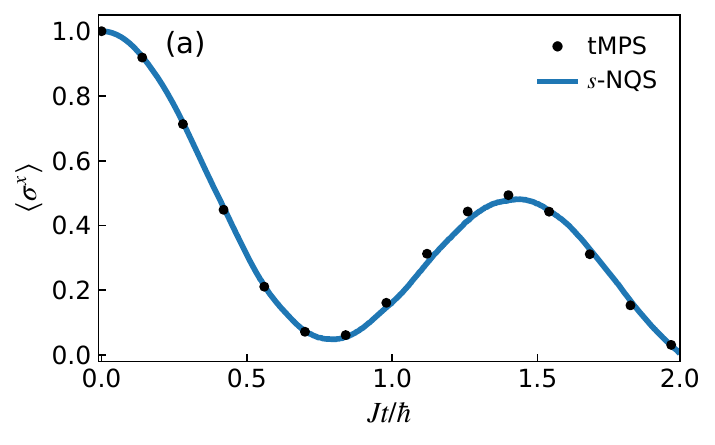}
    \caption{
        \R{
        Time evolution of the transverse magnetization $ \langle \sigma^{x} \rangle $ after quench in 1D tilted Ising model with $L = 40$ spins and open boundary conditions. The solid blue line represents the results from \textit{s}-NQS with $\alpha = 5$ and $Q = 5$, while the black dots correspond to time-dependent matrix product states (tMPS) simulations. The parameters are $J \Delta t/\hbar = 2.5 \times 10^{ - 4}$, $J \tau/\hbar = 0.2$, and $JT/\hbar = 2$.} }
    \label{fig:4}
\end{figure}

\R{To further test the scalability of the \textit{s}-NQS method, we perform simulations on a larger system of $L = 40$ spins with open boundary conditions. Fig.~\ref{fig:4} shows the evolution of the transverse magnetization $\langle \sigma^x \rangle$ obtained using \textit{s}-NQS compared against tMPS. The simulation uses the same ansatz size ($\alpha = 5$, $Q = 5$) as in smaller systems, along with a fine-grained time step $J\Delta t/\hbar = 2.5 \times 10^{-4}$ obtained via successive training refinements starting from $J\Delta t/\hbar = 5 \times 10^{-3}$. The results show close agreement with tMPS across the time window, demonstrating the robustness and scalability of \textit{s}-NQS toward larger systems. Notably, all results are obtained using a fixed Chebyshev expansion order $Q = 5$, showing that the basis remains expressive at increased time resolution and system size.}

\R{While the above results demonstrate the robustness of the method, the current implementation of \textit{s}-NQS also exhibits several technical limitations that may affect its performance in more demanding settings. 
In particular, the performance of \textit{s}-NQS appears highly sensitive to the choice of time step $\Delta t$, possibly due to the limitations of the second-order Taylor expansion used for time propagation. 
Such sensitivity may stem from the accumulation of truncation and non-unitarity errors, which can interfere with global optimization under smooth parameterization, where time-independent variational parameters must encode the dynamics across all time steps.
Additionally, although the compact ansatz remains effective at $L=40$, further increases in system size could require enhanced expressivity. 
These limitations, however, are not intrinsic to the \textit{s}-NQS framework. In future implementations, they could be mitigated by incorporating more robust integration schemes (e.g., Taylor-root or Padé product expansions~\cite{2024NC, GravinaVicentini2024}) or by enhancing the expressivity of the neural network ansatz.
}

\section{Conclusions}\label{sec:conclusions}  

In this work, we have introduced a smoothly parameterized neural quantum state ($s$-NQS) ansatz, in which the variational parameters are defined as linear combinations of temporal basis functions with time-independent variational parameters. Our approach achieves accurate simulations of the dynamics of a non-integrable many-body system using fewer cumulative parameters than step-by-step variational methods. Moreover, the smooth parameterization provides reliable initialization across intervals, leading to enhanced stability and efficiency in training.

To further enhance the effectiveness of the \textit{s}-NQS approach, future directions include using either alternative or higher-order approximations of the evolution operator, as proposed in Ref.~\cite{2024NC,GravinaVicentini2024}, and achieving this ansatz with more expressive network architectures such as convolutional neural networks or transformers\R{, which have already been adapted for variational wave functions; see Refs~\cite{schmitt2020quantum,gutierrez2022real,SchmittZurek2022,GravinaVicentini2024,vandewalle2024manybodydynamicsexplicitlytimedependent,2025arXiv250310462C,2021arXiv210802200W} for concrete implementations.}
Another promising direction involves exploring alternative temporal basis functions beyond the Chebyshev polynomials used in this work. Fourier modes, for instance, as employed in Ref.~\cite{sinibaldi2024timedependentneuralgalerkinmethod}, may offer advantages in capturing long-time behavior and periodic features of the system. Furthermore, while we employed the AdamW optimizer throughout this study, alternative optimization schemes, such as stochastic reconfiguration, may provide improved convergence properties and should be considered in future work. 
\R{As neural networks are not limited by system dimensionality, an important future direction is the application of \textit{s}-NQS to higher-dimensional quantum systems. In this work, we have focused on 1D systems due to the availability of accurate tMPS benchmarks and the use of a second-order Taylor expansion for time propagation.
Future versions of the framework may incorporate more scalable propagation schemes such as product-based approximations to address the increased computational complexity in higher-dimensional geometries.}
A thorough comparison with state-of-the-art neural quantum states methods as described in \cite{GravinaVicentini2024, sinibaldi2024timedependentneuralgalerkinmethod, vandewalle2024manybodydynamicsexplicitlytimedependent}, and potential integration with them, would also be an interesting research direction.   

An important underlying assumption of the \textit{s}-NQS ansatz is the continuity of the quantum evolution. A natural question we will tackle in future works is whether \textit{s}-NQS can effectively capture dynamics in systems undergoing dynamical phase transitions~\cite{heyl2018dynamical}, where certain quantities manifest a discontinuous behavior. We emphasize that true phase transitions occur only in the thermodynamic limit, making it essential to understand how the expressivity of the ansatz, especially the number and type of temporal basis functions, scales with system size. Beyond unitary time evolution, the \textit{s}-NQS approach could also be extended to non-unitary dynamics in open quantum systems, imaginary-time evolution for ground-state searches, and the study of classical systems.

\R{Our code is made publicly available at Ref.~\cite{snqs2025}.}

\section*{Acknowledgments}\label{sec:acknowledgments}  
We are grateful to A. Burger, E. Chong, A. Sinibaldi and L. Wang for fruitful discussions. D.P. acknowledges support from the Ministry of Education Singapore, under the grant MOE-T2EP50123-0046, 
D.W. acknowledges support from the National Research Foundation, Singapore and the Agency for Science, Technology and Research (A*STAR) under the Quantum Engineering Programme (NRF2021-QEP2-02-P03) and from CQT Core Funding Grant CQT\_SUTD\_2025\_01. W.Z acknowledges support from CQT Core Funding Grant A-0009609-38-00. 

The computational work for this article was partially performed at the National Supercomputing Centre, Singapore \cite{nscc}, and on cloud resources provided by NVIDIA Academic Grant Program. 

\bibliography{Bibliography.bib}

\appendix
\onecolumngrid

\section*{Appendix: Supplementary Material}

\section{Connecting two consecutive time windows}\label{app:window}
The first few Chebyshev polynomials and their first few derivatives are given by
\begin{align*}
\begin{array}{cccccc}
    &T_1(t) =  1, &T_{1}(t)' =  0 , \quad &T_1(t)''= 0 \quad &T_1(t)'''= 0  & \cdots\\
    &T_2(t) =  t, &T_{2}(t)' =  1 , \quad &T_2(t)''= 0 \quad &T_2(t)'''= 0  & \cdots\\
    &T_3(t) =  2t^{2}- 1,   &T_{3}(t)' =  4t,\quad  &T_3(t)''= 4 \quad &T_3(t)'''= 0  & \cdots\\ 
    &T_4(t) =  4t^{3}- 3t,   &T_{4}(t)' =  12t^2-3,\quad  &T_4(t)''= 24t \quad &T_4(t)'''= 24  & \cdots\\ 
    &\cdots & \cdots & \cdots  & \cdots & \cdots \\
\end{array}
\end{align*}

In our study, we consider the scaling-shifting functions $r_i(t)$ as simply a linear shift function with no scaling. To lighten the notation, in this section we will use $\theta^{[(k-1)T,\; kT]}_{j,q}\equiv \theta_{j,q}$ and $\theta^{[kT,\; (k+1)T]}_{j,q}\equiv \Theta_{j,q}$. We can thus write 
\begin{align}
     &\sum_{q = 0}^{Q-1}\theta_{j,q}  \left.\frac{d^{n} T_q(t)}{d t^{n}}\right|_{t=1} = 
 \sum_{q' = 0}^{Q-1} \Theta_{j,q'} \left.\frac{d^{n} T_q(t)}{d t^{n}}\right|_{t=-1}.       
\end{align}

For clarity, we here introduce an example. 
Taking $Q=3$, we can connect the windows from the second derivative  
\begin{align*}
    T_3(1)'' \theta_{j,3} &=  T_{3}(-1)'' \Theta_{j,3} \\
    \theta_{j,3} &=  \Theta_{j,3}   
\end{align*}
and then, obtained $\Theta_{j,3}$, we can, using the equations with the first time derivative, compute $\Theta_{j,2}$
\begin{align*}
    T_2(1)' \theta_{j,2} + T_3(1)' \theta_{j,3} &= T_{2}(-1)' \Theta_{j,2} + T_{3}(-1)' \Theta_{j,3} \\
    \theta_{j,2} + 4\theta_{j,3} &=  \Theta_{j,2} - 4\Theta_{j,3} \\
    \Theta_{j,2} &=  \theta_{j,2} + 8 \theta_{j,3},      
\end{align*} 
and we can thus compute $\Theta_{j,1}$ from             
\begin{align*}
    T_1(1) \theta_{j,1} + T_2(1) \theta_{j,2} + T_3(1) \theta_{j,3} &= T_{1}(-1) \Theta_{j,1} + T_{2}(-1) \Theta_{j,2} + T_{3}(-1) \Theta_{j,3} \\
    \theta_{j,1} + \theta_{j,2} + \theta_{j,3} &=  \Theta_{j,1} - \Theta_{j,2} + \Theta_{j,3} \\
    \Theta_{j,1} &= \theta_{j,1} + \theta_{j,2} + \theta_{j,3} + \theta_{j,2} + 8\theta_{j,3} - \theta_{j,3} \\
    \Theta_{j,1} &=  \theta_{j,1} + 2 \theta_{j,2} + 8 \theta_{j,3}. 
\end{align*}

\section{Taylor Expansion}\label{app:taylor}
Here, we give more details on how to use the Taylor expansion to compute the overlap $ C_{k} $ at time point $ k \Delta t $. The unitary operator is $U(\Delta t) = e^{-i H \Delta t}$, which can be approximated by Taylor expansion up to the second order        
\begin{equation}
    \begin{aligned}
        \hat{U}(\Delta t) &= e^{-i \hat{H} \Delta t} \approx \hat{\mathbb{I}} - \im \hat{H} \Delta t- \frac{1}{2}(\Delta t \hat{H})^2 \equiv  \hat{U}_\text{Taylor} 
    \end{aligned}
\end{equation}
where $\hat{\mathbb{I}}$ is the identity operator. 
Given an overlap $C_k$ at the time point $k\Delta t$ and $k\ne 0$, the overlap can be written as 
\begin{equation}
    \begin{aligned}
        C_k 
        &= 
        \frac{\langle\psi_{\mathbb{\vartheta}(k\Delta t)} |\hat{U}_\text{Taylor} |\psi_{\vartheta((k - 1)\Delta t)}\rangle}{\langle\psi_{\vartheta(k\Delta t)}|\psi_{\vartheta(k\Delta t)}\rangle}
        \cdot
        \frac{\langle\psi_{\vartheta((k - 1)\Delta t)}|\hat{U}_\text{Taylor}^{\dagger} |\psi_{\vartheta(k\Delta t)}\rangle} {\langle\psi_{\vartheta((k - 1)\Delta t)}|\psi_{\vartheta((k - 1)\Delta t)}\rangle} \\
        &= 
        C_{k - 1,k}(\hat{U}_\text{Taylor} ) \cdot C_{k,k - 1}(\hat{U}_\text{Taylor}^{\dagger} )
    \end{aligned}
\end{equation}
For the term $C_{k - 1,k}(\hat{U}_\text{Taylor} )$, we have 
\begin{equation}
    \begin{aligned}
        C_{k - 1,k}(U_\text{Taylor} ) 
        &=
        \frac{\langle \psi_{\vartheta(k\Delta t)} |\hat{\mathbb{I}} - \im \hat{H}\Delta t - \frac{1}{2}(\hat{H}\Delta t)^{2}| \psi_{\vartheta((k - 1)\Delta t)} \rangle}{\langle \psi_{\vartheta(k\Delta t)} |\psi_{\vartheta(k\Delta t)}\rangle} \\
        &= 
        \frac{\langle \psi_{\vartheta(k\Delta t)} |\hat{\mathbb{I}}| \psi_{\vartheta((k - 1)\Delta t)} \rangle}{\langle \psi_{\vartheta(k\Delta t)}| \psi_{\vartheta(k\Delta t)}\rangle} 
        - 
        \mathrm{i} \Delta t\frac{\langle \psi_{\vartheta(k\Delta t)}| \hat{H} | \psi_{\vartheta((k - 1)\Delta t)} \rangle}{\langle \psi_{\vartheta(k\Delta t)} |\psi_{\vartheta(k\Delta t)}\rangle}  
        - 
        \frac{1}{2} (\Delta t)^{2}\frac{\langle\psi_{\vartheta(k\Delta t)}|\hat{H} \hat{H}|\psi_{\vartheta((k - 1)\Delta t)}\rangle}{\langle\psi_{\vartheta(k\Delta t)}|\psi_{\vartheta(k\Delta t)}\rangle}
    \end{aligned}\label{eq:Ckk}
\end{equation}
We can rewrite the first term of Eq.~(\ref{eq:Ckk}) as      
\begin{equation}
    \begin{aligned}
        \frac{\langle\psi_{\vartheta(k\Delta t)}|\hat{\mathbb{I}}|\psi_{\vartheta((k - 1)\Delta t)}\rangle}{\langle\psi_{\vartheta(k\Delta t)}|\psi_{\vartheta(k\Delta t)}\rangle} 
        &= 
        \frac{\sum_{\boldsymbol{x}}\psi_{k\Delta t}^*(\boldsymbol{x}) \sum_{\boldsymbol{x}'}^{} \mathbb{I}_{\boldsymbol{xx'}} \psi_{(k - 1)\Delta t}(\boldsymbol{x}')}{\sum_{\boldsymbol{\omega}}|\psi_{k\Delta t}(\boldsymbol{\omega})|^2} \\
        &= 
        \sum_{\pmb{x}}\frac{|\psi_{k\Delta t}(\boldsymbol{x})|^2 }{\sum_{\boldsymbol{\omega}}|\psi_{k\Delta t}(\pmb{\boldsymbol{\omega}})|^2}\sum_{\boldsymbol{x}'}^{} \mathbb{I}_{\boldsymbol{x}\boldsymbol{x'}} \frac{\psi_{(k - 1)\Delta t}(\boldsymbol{x}')}{\psi_{k\Delta t}(\boldsymbol{x})}\\
        &= 
        \sum_{\boldsymbol{x}}^{} P_{k\Delta t}(\boldsymbol{x}) \frac{\psi_{(k - 1)\Delta t}(\boldsymbol{x})}{\psi_{k\Delta t}(\boldsymbol{x})},  
    \end{aligned}
\end{equation}
the second term as 
\begin{equation}
    \begin{aligned}
        \frac{\langle\psi_{\vartheta(k\Delta t)}|\hat{H}|\psi_{\vartheta((k - 1)\Delta t)}\rangle}{\langle\psi_{\vartheta(k\Delta t)}|\psi_{\vartheta(k\Delta t)}\rangle} 
        &= 
        \frac{\sum_{\boldsymbol{x}}\psi_{k\Delta t}^*(\boldsymbol{x}) \sum_{\boldsymbol{x}'}^{} H_{\boldsymbol{xx'}} \psi_{(k - 1)\Delta t}(\boldsymbol{x}')}{\sum_{\boldsymbol{\omega}}|\psi_{k\Delta t}(\boldsymbol{\omega})|^2} \\
        &= 
        \sum_{\pmb{x}}\frac{|\psi_{k\Delta t}(\boldsymbol{x})|^2 }{\sum_{\boldsymbol{\omega}}|\psi_{k\Delta t}(\pmb{\boldsymbol{\omega}})|^2}\sum_{\boldsymbol{x}'}^{} H_{\boldsymbol{x}\boldsymbol{x'}} \frac{\psi_{(k - 1)\Delta t}(\boldsymbol{x}')}{\psi_{k\Delta t}(\boldsymbol{x})}\\
        &= 
        \sum_{\boldsymbol{x}}^{} P_{k\Delta t}(\boldsymbol{x}) E_\text{loc}(\boldsymbol{x}),  
    \end{aligned}
\end{equation}
and the third term as     
\begin{equation}
    \begin{aligned}
        \frac{\langle\psi_{\vartheta(k\Delta t)}|\hat{H} \hat{H}|\psi_{\vartheta((k - 1)\Delta t)}\rangle}{\langle\psi_{\vartheta(k\Delta t)}|\psi_{\vartheta(k\Delta t)}\rangle} 
        &= 
        \frac{\langle\psi_{\vartheta(k\Delta t)}|\hat{H} \sum_{\boldsymbol{y}}\vert \boldsymbol{y} \rangle\langle \boldsymbol{y} \vert   \hat{H}|\psi_{\vartheta((k - 1)\Delta t)}\rangle}{\langle \psi_{\vartheta(k\Delta t)}|\psi_{\vartheta(k\Delta t)}\rangle} \\
        &= 
        \frac{\sum_{\boldsymbol{x}}\psi_{k\Delta t}^*(\boldsymbol{x}) \sum_{\boldsymbol{y}}H_{\boldsymbol{x}\boldsymbol{y}} \sum_{\boldsymbol{x}'}^{} H_{\boldsymbol{yx'}} \psi_{(k - 1)\Delta t}(\boldsymbol{x}')}{\sum_{\boldsymbol{\omega}}|\psi_{\Delta kt}(\boldsymbol{\omega})|^2} \\
        &= 
        \sum_{\pmb{x}}\frac{|\psi_{k\Delta t}(\boldsymbol{x})|^2 }{\sum_{\boldsymbol{\omega}}|\psi_{k\Delta t}(\pmb{\boldsymbol{\omega}})|^2} \sum_{\boldsymbol{y}}H_{\boldsymbol{x}\boldsymbol{y}} \sum_{\boldsymbol{x}'} H_{\boldsymbol{y}\boldsymbol{x'}} \frac{\psi_{(k - 1)\Delta t}(\boldsymbol{x}')}{\psi_{k\Delta t}(\boldsymbol{x})}\\
        &= 
        \sum_{\boldsymbol{x}} P_{kt}(\boldsymbol{x}) \sum_{\boldsymbol{y}}H_{\boldsymbol{x}\boldsymbol{y}} \sum_{\boldsymbol{x}'} H_{\boldsymbol{y}\boldsymbol{x'}} \frac{\psi_{(k - 1)\Delta t}(\boldsymbol{x}')}{\psi_{k\Delta t}(\boldsymbol{x})}. 
    \end{aligned}
\end{equation}

\section{System-size dependence and time step effects}\label{app:scaling}

\R{To examine how the performance of the \textit{s}-NQS approach depends on system size and time steps, we evaluate the converged training loss $\log(\mathcal{C})$ as a function of system size $L$ for several choices of the time step $\Delta t$.}

\R{Specifically, we consider system sizes from $L=10$ to $L=30$, using fixed ansatz parameters $\alpha = 5$ and $Q = 5$. All simulations are performed with $5{,}000$ Monte Carlo samples. The reported loss values correspond to the average over the final $100$ epochs of a $4000$-epoch training run, after convergence has been reached.}

\begin{figure}[htbp]
    \centering
    \includegraphics[width=0.55\linewidth]{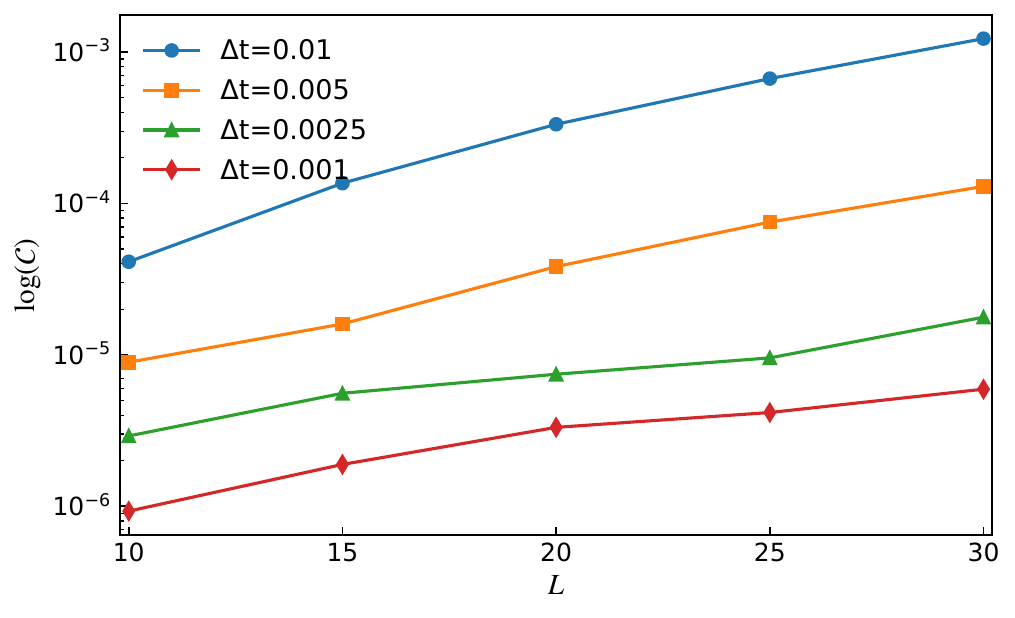}
    \caption{
    \R{System-size dependence of the converged training loss $\log(\mathcal{C})$ in the first time interval, for different choices of time step $\Delta t$. All simulations use the same ansatz parameters ($\alpha = 5$, $Q = 5$) and $5{,}000$ Monte Carlo samples. Each point is averaged over the final $100$ epochs of a $4000$-epoch training run.}
    }\label{fig:scaling_app}
\end{figure}

\R{The results in Fig.~\ref{fig:scaling_app} show that the training loss increases with system size for fixed $\Delta t$, reflecting the growing complexity of the wavefunction. At the same time, smaller values of $\Delta t$ consistently yield lower loss values across all system sizes, underscoring the importance of fine temporal resolution for stable and accurate training. We expect that this trend is not fundamental to the \textit{s}-NQS architecture, but rather linked to the chosen expansion scheme. Using higher‑order product‑form propagation methods should reduce accumulated error at fixed $\Delta t$, enabling more favorable scaling with system size. Exploring such integrators is therefore a natural and promising direction for future large‑scale applications.}

\end{document}